# Radon mitigation by soil depressurisation case study: radon concentration and pressure field extension monitoring in a pilot house in Spain

Marta Fuente[1,2,3], Jamie Goggins[2,3], Daniel Rábago[4], Ismael Fuente[4], Carlos Sainz[4], Mark Foley[1*]

[1] School of Physics, National University of Ireland Galway, Ireland

[2] Civil Engineering, College of Engineering & Informatics, National University of Ireland Galway, Ireland

[3] Centre for Marine and Renewable Energy (MaREI), Ryan Institute, National University of Ireland Galway, Ireland

[4] Radon group, University of Cantabria, Spain

* Corresponding author

E-mail address: mark.foley@nuigalway.ie, m.fuentelastra1@nuigalway.ie

**Abstract**

A one-year monitoring study was conducted in a pilot house with high radon levels to investigate the ability and efficiency of radon mitigation by soil depressurisation (SD) both active and passive. The study included monitoring of radon concentration, pressure field extension (pfe) under the slab and some atmospheric parameters for different testing phases. Periods in which the house remained closed to foster radon accumulation were alternated with phases of active and passive soil depressurisation under different conditions. The behaviour of the radon concentration in the pilot house was analysed along with the influence of atmospheric variables, significant correlations were found for the radon concentration with atmospheric pressure, outdoor temperature and wind. From the pfe analysis it was proven that the pressure drop with distance from the suction point of the SD system is proportional to the depressurisation generated. It was found also that the permeability characterisation of the pilot house agrees with the literature about granular fill materials characterisation for radon SD systems across Europe. Radon reductions in excess of 85% were achieved for the different testing phases in all cases. Finally, from the results it was stated that a fan power of 23 W is sufficient to ensure radon reductions over 85% for dwellings with similar aggregate layer and soil permeability.

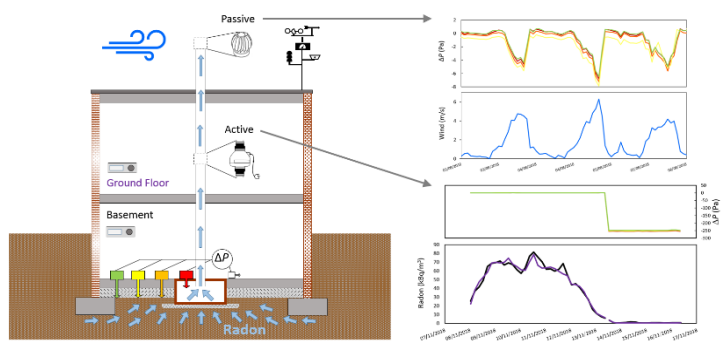

## 1. Introduction

Radon ($^{222}$Rn) is a colourless, odourless, radioactive gas formed in the ground by the radioactive decay of uranium ($^{238}$U), which is present in all rocks and soils of the Earth's crust. With a half-life of 3.8 days together with its noble gas condition, radon can move through interconnected pores in the soil, reach the Earth's surface and penetrate into buildings. Radon is the greatest natural source of exposure to ionising radiation for the general public and it is also the leading cause of lung cancer after smoking, as stated by the World Health Organisation (WHO, 2009). Poor ventilation conditions, gaps or cracks in the construction systems favour the accumulation of radon inside buildings, leading to health risks related to the inhalation of the radon decay products. A 9% of deaths from lung cancer per year are attributable to residential radon exposure in the European Union, which accounts for more than 20,000 deaths each year (Darby et al., 2004; WHO, 2018).

There are various prevention and mitigation measures that might be considered to minimise indoor radon concentration, in order to address the radon problem both in new and existing buildings. Radon protection strategies include reduction of radon entry by sealing of surfaces, barriers or membranes, soil depressurisation (SD) techniques to reverse the air pressure differences between the indoor occupied space and the soil underneath the building, and ventilation of spaces to dilute indoor radon concentration with external air (Long et al., 2013; Jiránek, 2014; WHO, 2018).

The active and passive SD techniques have proven to be the most effective strategy for indoor radon prevention and mitigation. SD systems include three basic components: a suction point, ideally located in a continuous and uniform permeable aggregate layer under the slab, an exhaust pipe to extract the soil gas and a means of extraction, which can be a mechanical fan in case of forced extraction or a cowl for passive depressurisation using the wind force. The suction point is normally a sump placed under the slab or on a side of the building, connected to a permeable aggregate layer, but perforated pipes beneath the existing floors can be an alternative to sumps (DELG, 2002; Abdelouhab et al., 2010; Long et al., 2013).

Previous works discuss the importance of the aggregate layer in the design of SD systems for radon mitigation. The impact of the granular fill materials permeability of such aggregate layer and the soil permeability beneath and surrounding the building on the SD effectiveness has been investigated and permeability characterisation of aggregates within the European context conducted (Hung et al, 2018a; 2018b; Fuente et al., 2019). But there is a lack of evidence in testing efficiency of SD techniques in relation with the pressure distribution in actual buildings with elevated radon concentration where radon reduction can be quantified with confidence.

This paper outlines a case study on radon mitigation by soil depressurisation in a real building. A one-year monitoring study was conducted in a pilot house with very high radon levels to investigate the ability and efficiency of active and passive SD. The work includes the behaviour analysis of the radon concentration inside the experimental building. Also, it presents the analysis of the SD effectiveness, looking at the pressure distribution induced under the building slab, in relation to the permeability characterisation of the aggregate layer beneath the slab, and the achievable radon reduction in such conditions.

## 2. Materials and methods

*2.1 Pilot house: location and design*

The pilot house chosen for the case study is located in Saelices el Chico, Salamanca (Spain) within the land of a former uranium mine managed by the company ENUSA Industrias Avanzadas S.A. (see Figure 1) now under restoration activities. The location of the experimental house was selected due to the high radon exhalation rate and the high radium content in the soil of the area, which would provide high radon levels accumulated inside the building. An average radium concentration of 1600 Bq/kg was quantified from different soil samples taken onsite, this value is 40 times in excess of the average world-wide concentration, approximately 40 Bq/kg (Frutos et al., 2011).

The experimental house was designed to reproduce a space large enough to be representative of a room in a typical dwelling house. It consists of two storeys, a partially below grade so-called basement and a ground floor connected by a standard door. The dimensions of the rooms are 5 x 5 m$^2$ (see Figure 2). There are two windows at the ground floor level, one in the front wall next to the main door and another one in the opposite wall. The front wall of the house is facing the North.

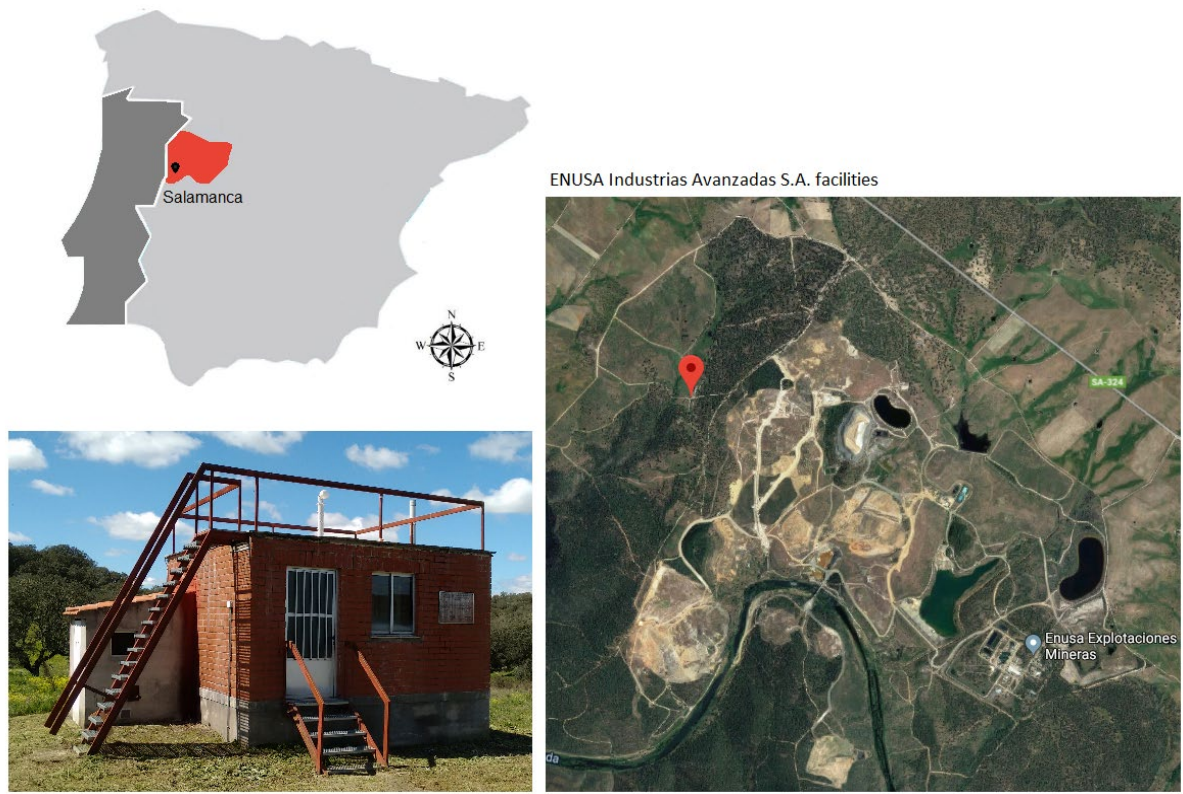

Figure 1. Map of Spain and plan of the mining facilities indicating the location of the pilot house and a recent picture of the building.

In 2006 when the pilot house was built for a different study, several mitigation measures were investigated (Frutos 2009). As a result, there are two soil depressurisation systems installed in the house. Both SD systems consist of a 1 m² and 0.5 m deep sump and an exhausting pipe, one system is located in the centre of the experimental house with the sump placed in the aggregate layer below the concrete slab and the other system is placed on a side of the house (see Figure 2).

Materials used for the construction of the house were according the Spanish building practices, a 15 cm thick aggregate layer was placed below a 10 cm thick concrete slab. Standard clay bricks were used for the walls and conventional perforated clay bricks to build the sumps of the SD systems (Frutos et al. 2011).

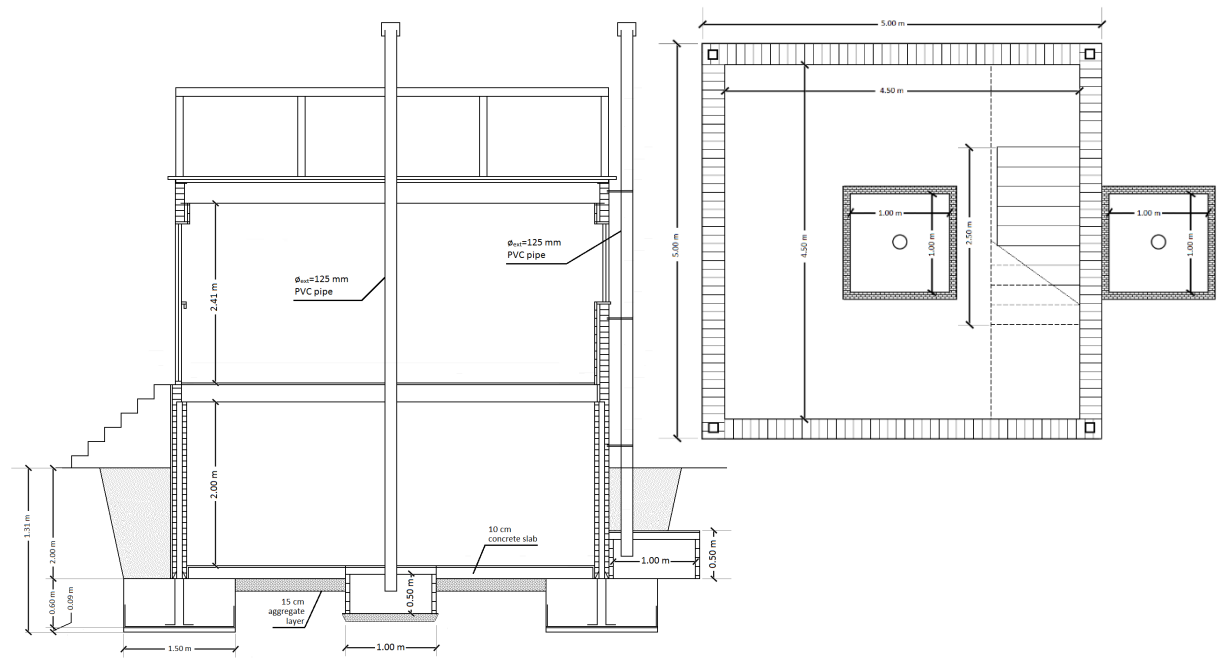

Figure 2. Section view of the house and plan of the basement after installation of mitigation measures, modified from Frutos et al. (2011).

*2.2 Monitoring system*

To continuously monitor radon concentration several active radon monitors were used, including the Radon Scout (SARAD GmbH), Radon Scout Home (SARAD GmbH) and AlphaE (Bertin Instruments) detectors. Performance of the radon monitors used in the experiment at the pilot house was tested previously in a purpose-built radon chamber (Fuente et al., 2018). Radon concentration was recorded in the basement and in the ground floor.

For some measurement periods, the radon monitors used were contaminated due to the high radon exposure levels at the pilot house. These monitors were then replaced. There were also some problems due to memory of the devices in some cases, so there is radon data missing for some of the testing phases.

A pressure sensor system was installed to monitor the distribution of pressure under the slab of the house. The pressure system was specifically developed for this experiment at the pilot house in collaboration with a research group of the ITEFI-CSIC, Madrid (Spain). It is an acquisition system designed with segmented architecture and capacity up to 15 pressure sensors. It consists of an adaptor board for the pressure sensor units and contains a series of Honeywell pressure sensors with SPI communication. The connections between the units use Ethernet cables connected in parallel and the adaptor board needs to be connected to a PC by an input/output USB card type Lab Jack U3. The actual system installed in the house for this experiment consists of a total of 8 pressure sensor units. There are 5 of them distributed along the basement area in different holes drilled through the concrete slab to measure pressure difference in the aggregate layer under the slab and the inhabited volume of the basement, at distances $d$= 1, 2 and 2.4 m from the central sump. The remaining 3 pressure sensors are placed at the sump and pipe of the central SD system and at room level for reference.

To record atmospheric conditions locally at the house site, a local weather station (PCE-FWS20, PCE Instruments) was installed on the rooftop. Variables recorded were wind velocity, outdoor temperature, atmospheric pressure, relative humidity percentage and accumulated rain. Both the pressure sensor system and the weather station are remotely accessible which facilitates data collection.

*2.3 Experimental methodology*

The initial monitoring plan was alternating testing phases of SD performance (active or passive) with periods in which the house remained completely closed, in order to record radon increase and reduction over the different phases along with the pressure field extension induced under the slab, hereinafter referred as pfe.

The monitoring study commenced in June 2018 with a first phase of the house closed to foster accumulation of radon gas. All testing phases with this setting, in which the house remains closed and the pipes of the SD systems capped to foster radon accumulation in the building, will be henceforth referred as closed periods. After the first closed period (phase 1), a phase 2 involved passive SD performance evaluation. Then a subsequent series of closed periods followed by active SD performance was conducted up to 9 phases, with different active SD settings, ending in April 2019.

Only the central SD system was used for the investigation of the soil depressurisation during the SD testing phases. A rotating cowl was used for the passive SD operation and for the active SD performance, a mechanical fan (RP145i, RadonAway with 80 W max) was installed in the central SD system pipe. The mechanical fan was modified by adding a potentiometer to control the extraction airflow, which in terms of velocity ranges from 0 to 3.5 m/s. A schema of the experimental house settings for the different phases is shown in Figure 3.

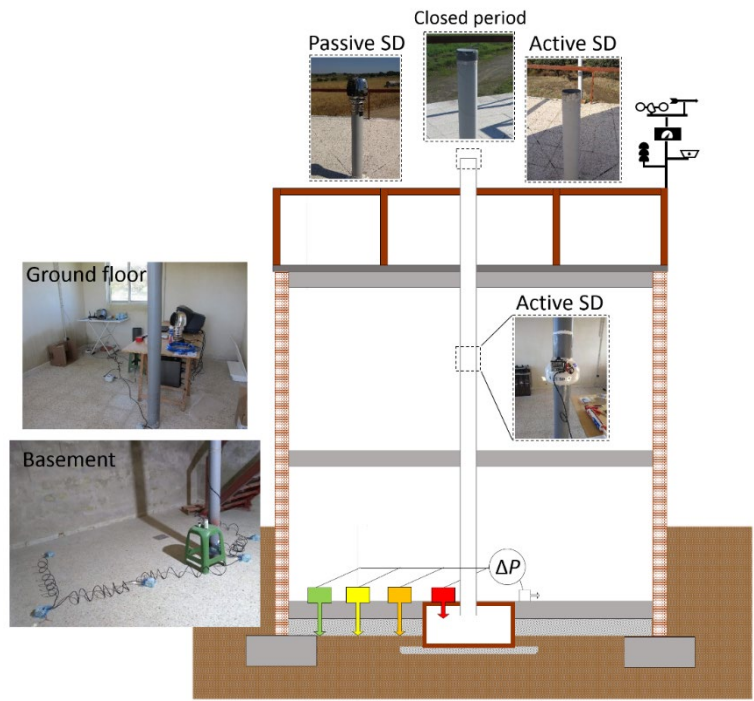

Figure 3. Schema of the pilot house for the closed periods, active SD and passive SD testing phases. Section view of the experimental house shows the pressure sensors system in the basement (same for all phases), the mechanical fan installed on the central pipe (for the active SD performance) and the top of the exhaust pipe in the house rooftop, with a cap, a rotating cowl or opened, depending on the testing phase.

The duration of the different phases varied, depending on the access to the site and technical problems experienced with the sensors or the power supply. The monitoring study was stopped at some times and later resumed.

From February 2019, a stage of the monitoring study focused on the investigation of the mechanical fan extraction impact on the SD effectiveness was conducted. It consisted of short periods (1-2 weeks) of active SD followed by closed periods, gradually increasing the mechanical fan extraction for the SD performance by controlling the airflow rate.

A summary of the testing phases including dates and incidents is presented in Table 1.

Table 1. Summary of testing phases at the pilot house.

| Phase | Description | Dates |
|---|---|---|
| 1 | Closed period | 25/06/2018 – 26/07/2018 |
| 2 | Passive SD | 26/07/2018 – end of August |
| Stopped period (issues related with the pressure sensors) | | |
| 3 | Closed period | 10/10/2018 – 13/11/2018 |
| 4 | Active SD ($v_{ext}$= 0 – 3.5 m/s) | 13/11/2018 – 16/11/2018 |
| Stopped period (issues related to power supply in the house | | |
| 5 | Closed period | 13/12/2018 – 19/02/2019 |
| 6 | Active SD ($v_{ext}$= 1.5 m/s) | 19/02/2019 – 06/03/2019 |
| 7 | Closed period | 06/03/2019 – 14/03/2019 |
| 8 | Active SD ($v_{ext}$= 2 m/s) | 14/03/2019 – 02/04/2019 |
| 9 | Closed period | 02/04/2019 – 30/04/2019 |

Radon levels were monitored continuously, but also, passive radon detectors were used for some testing periods. However, due to the high radon concentration the passive track etched detectors were saturated in the most cases.

**3. Radon concentration behaviour in the pilot house**

There is a long term record of the radon concentration fluctuations in the experimental house measured during the different testing phases. Before looking at the radon reductions generated as a result of the soil depressurisation, it is important to try to understand the natural behaviour of the radon concentration inside the house. To do so, the closed testing periods when there were no mitigation measures in operation and the house remained closed with the exhaust pipes of the SD systems capped are analysed.

Indoor radon levels in the experimental house depend on three features: the radon source, the entry rate and the air exchange between the building and the outdoor air, all of which, in turn, depend on many other variables and especially atmospheric conditions.

The radon source is constituted by the soil beneath and surrounding the house, which contains high radium levels. Therefore, it is expected to find higher radon levels in the basement, which is partially below grade and in direct contact with the soil, than in the ground floor. An overview of the radon levels recorded is presented in Figure 4.

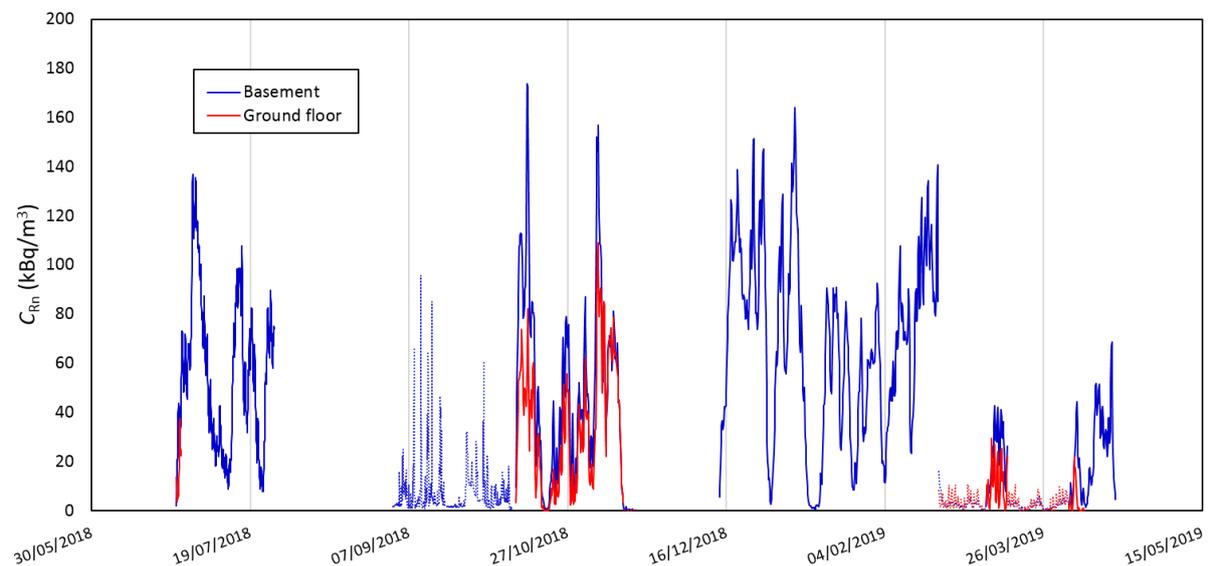

Figure 4. Radon concentration recorded in the basement and ground floor of the pilot house. The solid line indicates the closed testing periods and the dashed line indicates the periods of SD performance.

An average radon concentration of 55 kBq/m$^3$ is found in the basement for the closed house conditions, while for the ground floor there is an average radon concentration of 26 kBq/m$^3$ under the same house settings. Both values are obtained from the radon records available during the closed testing periods. The average radon concentration values from the initial study conducted in 2006 are 40 kBq/m$^3$ for the basement and 7 kBq/m$^3$ for the ground floor. These values were obtained from a three month measurement period (January-April) in which the experimental house remained closed building up the radon concentration, before the installation of any mitigation measures (Frutos et al., 2011). The differente may be related to the deterioration of the basement slab associated with thermal dilation or other analogous phenomenon, bringing as a consequence the opening of new crack or radon pathways.

The concentration ratio between floors found for the closed house testing configuration is approximately two, which means that the concentration recorded in the basement is approximately double the concentration in the ground floor. This result is according to expectation, as the basement is in direct contact with the soil and the main radon gas entry is through gaps or cracks in the foundation, while the radon in the ground floor comes from the radon in the basement. The infiltration through the main door and the windows could be considered as a source of radon gas, but the outdoor air radon concentration at the site is much lower, approximately 300 Bq/m$^3$. Then, assuming that radon in the ground floor comes only from the basement, the lower radon concentration in the ground floor is explained by the radioactive decay and the exchange of outdoor air through infiltration.

*3.1 Radon behaviour and atmospheric parameters*

Fluctuations of radon concentration are daily and seasonal, and they are related to atmospheric conditions and the air exchange between the building and the outdoor air. There is also a seasonal component related to the outdoor temperature changes and the associated atmospheric pressure variations that directly affects radon entry in the building (Nero et al., 1990; Scivyer et al., 1998).

The air exchange rate in the experimental house reaches its minimum value during the closed testing periods, as there is no ventilation mechanism. Thus, the radon levels in the house basically depend on the atmospheric conditions, which determine the soil gas pressure-driven flow from the ground into the building.

The differences between indoor and outdoor temperature in a building can generate a pressure gradient due to the Stack effect, leading to an increase of the soil gas flow from the ground into the basement through the existing entry routes. However, the pressure gradient created due to temperature differences is small compared to the caused by other parameters. The wind effect can generate a pressure gradient too, due to the pressure changes and suction generated in the walls, which can modify the indoor pressure relative to that in the ground. But also, the wind causes the opposite effect as it fosters ventilation through infiltration that lower the radon levels (Abdelouhab et al. 2010; Burke et al., 2010).

Only under certain weather conditions it is possible to find clear correlations between the trends in radon concentration and atmospheric variables, as multiple factors affect. The trends of the different atmospheric parameters recorded were analysed in relation to the radon concentration recorded in the pilot house, for the closed testing periods. Selected measurement periods where statistical significant correlation was found between the atmospheric variables and radon levels in the pilot house are presented in Figure 5.

Theoretically, pressure variation inside a dwelling is almost simultaneous with atmospheric pressure changes, but pressure changes in the soil pores beneath the building reach the atmospheric pressure values with a time delay that depends on the soil characteristics (e.g. porosity). Thus, there is a pressure gradient generated between the soil and inside the building that leads to an increase of the soil gas airflow into the building. From the analysis of the experimental data recorded, a negative correlation is found between the atmospheric pressure and radon levels, which is consistent with theory. Figure 5a, depicts radon concentration decreasing with the increase of the atmospheric pressure with a time delay, Pearson's correlation coefficient found for this case is $r = -0.44$ with a 95% confidence level.

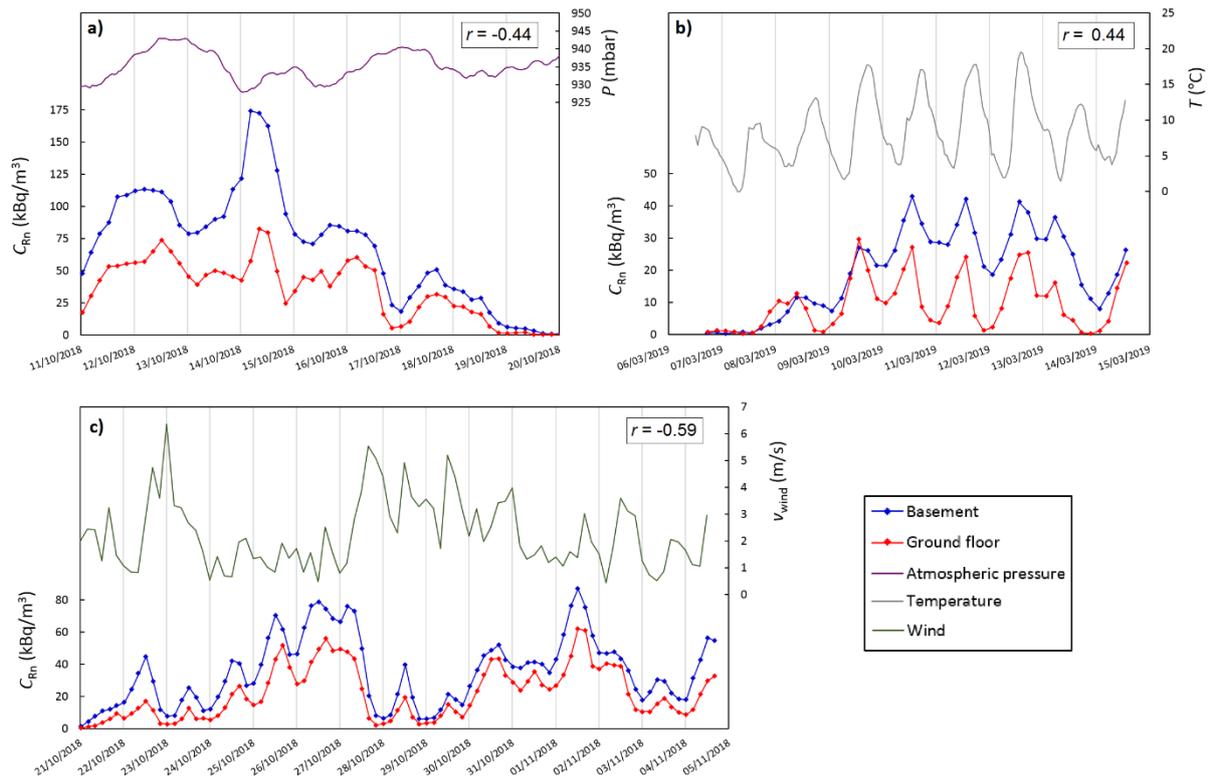

Figure 5. Radon concentration recorded in the basement and ground floor of the pilot house along with atmospheric variables, a) atmospheric pressure, b) outdoor temperature and c) wind velocity, for different time periods while the house was closed. Pearson's correlation coefficient between radon concentration in the basement and the corresponding atmospheric variable is indicated.

From the analysis of the data recorded at the experimental house, a positive correlation between outdoor temperature and radon concentration is found only at a daily scale. From Figure 5b data, it is obtained a Pearson's correlation coefficient of $r = 0.44$ with a 95% confidence level.

A negative correlation between the radon concentration and the wind is found, wind velocity increases when the radon level decreases as it is observed in Figure 5c. For this case an $r = -0.59$ Pearson's correlation coefficient is obtained with 95% confidence level. No significant correlation with radon levels was found for the other two atmospheric parameters monitored in the study, relative humidity percentage and accumulated rain.

Looking at diurnal radon fluctuations, apart from the difference in the radon concentration levels from the basement and ground floor, the radon concentration records are in some cases temporally shifted one from the other. The explanation for this fact, considering the concentration in the basement as the reference, is that the radon in the ground floor comes from the basement and the exchange between floors takes a few hours, causing a time delay in the radon concentration.

This case study was focused on the impact of the soil depressurisation, so further investigation extended in time would be required to understand the radon behaviour in detail at the pilot house, as a function of all the atmospheric parameters and their time variations.

## 4. Soil Depressurisation effectiveness analysis

The distribution of the pressure induced under the slab as a consequence of the soil depressurisation system performance and radon reduction are analysed below.

*4.1 Pressure Field Extension*

The pressure distribution under the pilot house was studied for different depressurisation induced at the central sump of the SD system, both by active and passive performance of the system. Prior to the depressurisation analysis, it was found that the pressure difference between the indoor air and the measurement points under the slab fluctuates around 0 Pa for the closed testing periods when there is no SD in operation. During the passive SD testing period, the pressure induced under the slab as a consequence of the wind force reached levels of -20 Pa, an example of the passive depressurisation in relation to the wind velocity is shown in Figure 6.

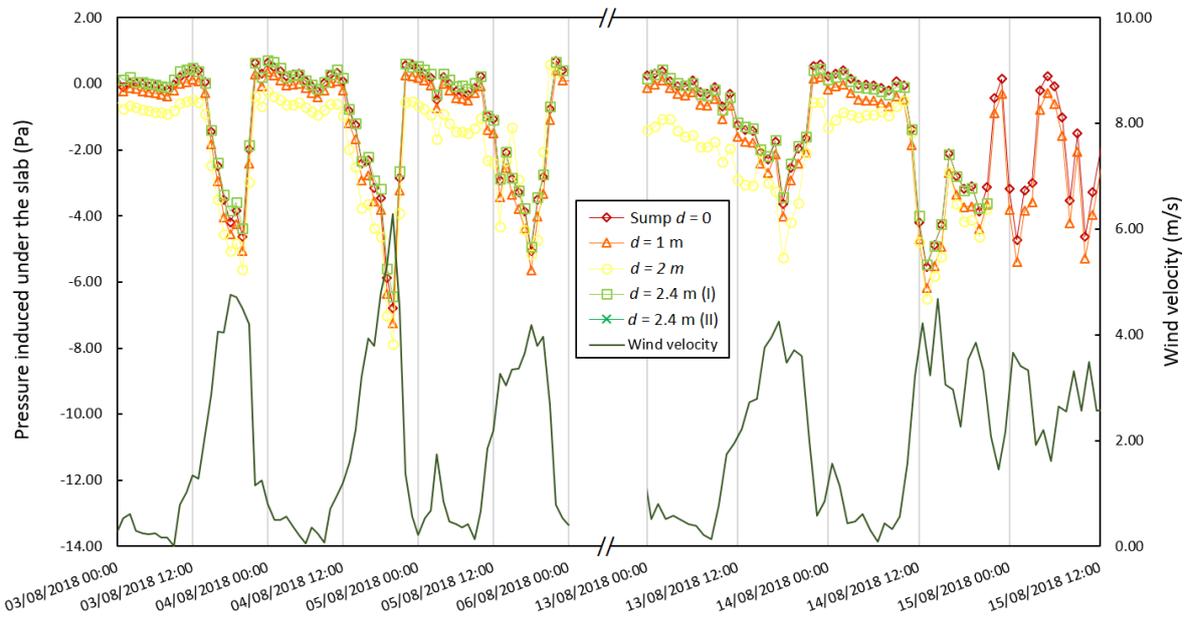

Figure 6. Hourly averaged pressure induced under the slab at the sump and the different measurement points with distances indicated from the central pipe along with hourly averaged wind velocity recorded at the site.

The analysis of the pressure data recorded under the slab for the different testing phases at distances $d$= 1, 2 and 2.4 m from the suction point, using the centre of the sump as the reference, leads to obtain the rate of pressure drop with distance across the slab, which is also related to the depressurisation generated (see Figure 7).

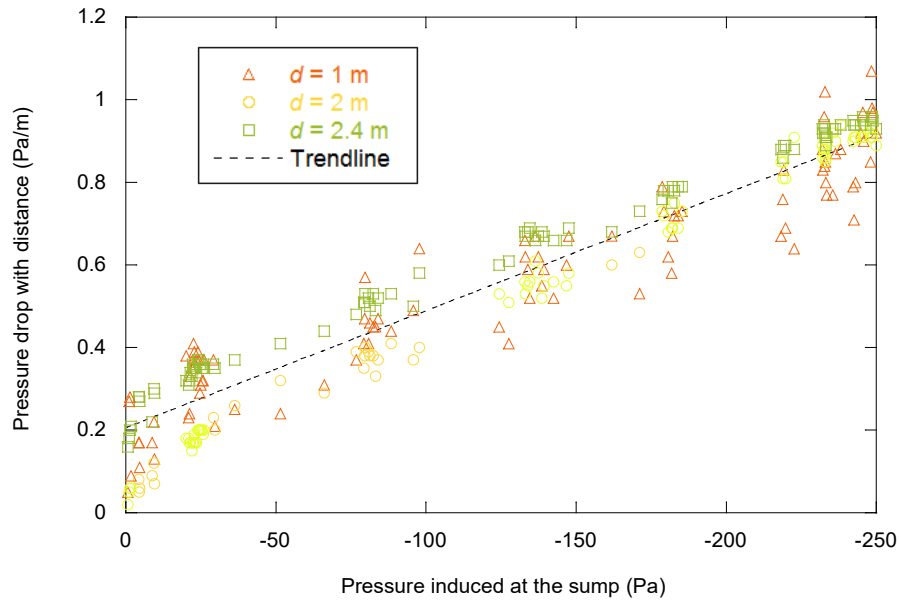

Figure 7. Pressure drop with distance against pressure induced at the sump. Dots represent the experimental data obtained from pressure records at distances $d$= 1, 2 and 2.4 m from the sump, both for active and passive SD testing phases. The dashed line is the linear trend obtained from all the experimental data, with a coefficient of determination $R^2$=0.95.

Figure 7 shows that the pressure drop with distance results are consistent for the measurements recorded at the different distances $d$= 1, 2 and 2.4 m from the suction point. The trend of the pressure drop with distance is linear with the depressurisation under the slab, therefore the lower the pressure induced under the slab, the higher the pressure drop with the distance. However, it was found a quite homogenous pfe, not exceeding 1 Pa/m pressure drop rate with distance for the highest depressurisation tested, induced by the highest extraction airflow rate permitted by the fan during the active SD operation.

Results of the pressure induced under the slab are presented in Figure 8 as a function of the extraction velocity in the exhaust pipe, for the active SD operation by means of the mechanical fan (a) and for passive SD operation by means of a rotating cowl (b).

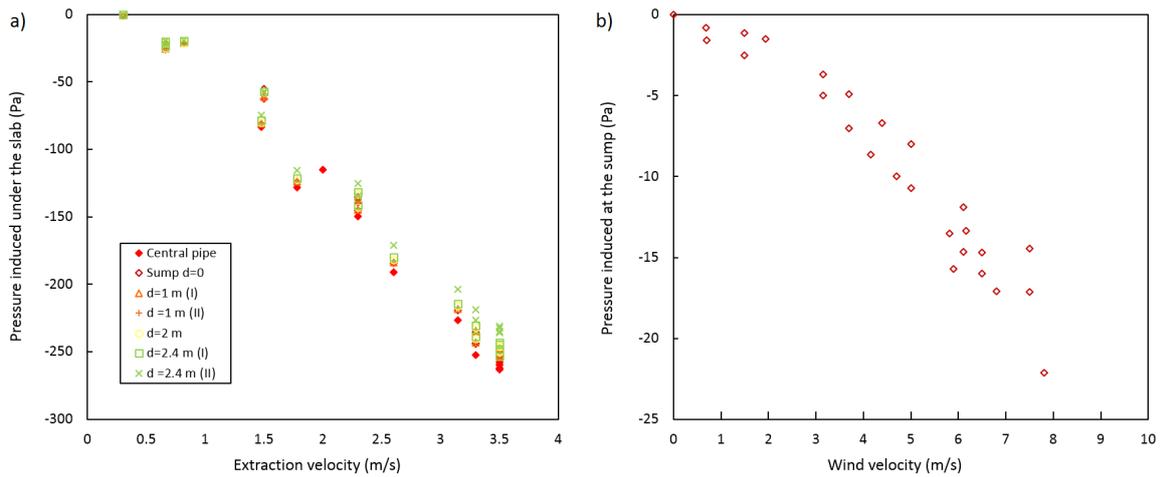

Figure 8. (a) Pressure induced under the slab against the extraction velocity of the mechanical fan for the different measurement points at different distances from the suction point. (b) Pressure induced at the sump under the slab against the wind velocity.

Results from the passive SD operation show a lower depressurisation induced at the sump by the wind velocity compared to depressurisation generated as a result of the active SD operation using a mechanical fan. The highest depressurisation under the slab recorded during the passive SD operation, induced by the wind force using a rotating cowl, is -22 Pa and it corresponds to wind velocities up to 8 m/s. However, the highest depressurisation recorded under the slab during the active SD operation is around -250 Pa, induced by the highest extraction airflow rate of the mechanical fan equivalent to 3.5 m/s.

Abdelouhab et al. (2010) conducted a study of this kind in France at the MARIA (Mechanized house for Advanced Research on Indoor Air) experimental house, built with a 40 cm thick aggregate layer beneath the slab and two sumps, one centred and another decentred placed on the aggregate layer. They calculated two behaviour laws to relate the extraction airflow $Q$ with the pressure difference induced between the aggregate layer and the inhabited volume $\Delta P$ for the natural and mechanical extraction. A similar behaviour law is obtained from the active SD experimental data of the monitoring study at the pilot house (see Figure 9) but in this case it is clear that the extraction airflow needed to generate the same pressure difference is lower compared to the MARIA house.

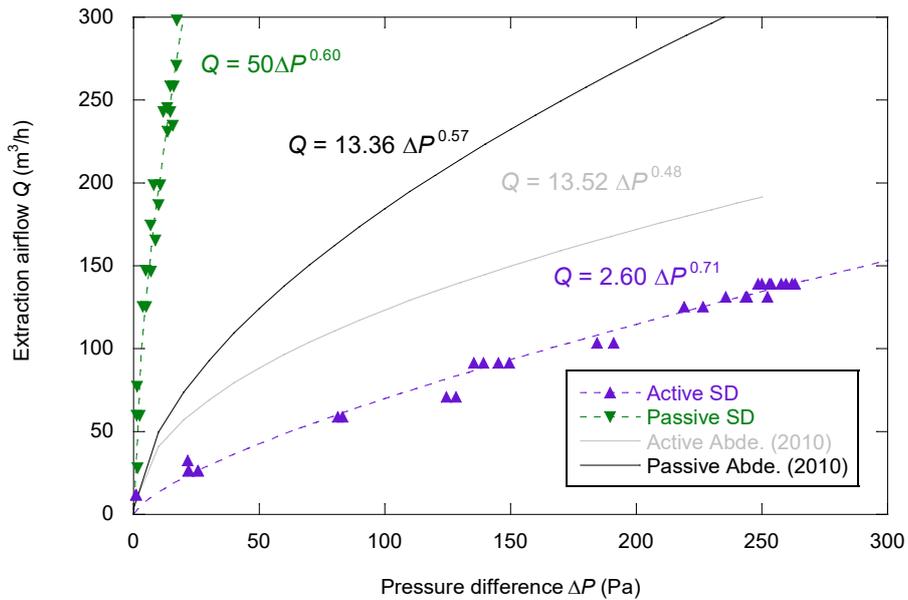

Figure 9. Extraction airflow through the central exhaust pipe as function of the pressure difference generated between the aggregate layer under the slab and the inhabited volume of the basement. Dots represent the experimental data for the active SD (purple) and passive SD (green), the dashed lines are the trends obtained from the experimental data and the solid lines represent the law relating extraction airflow with depressurisation for active and passive SD from Abdelouhab et al. (2010).

From Figure 9 it is observed that the behaviour law differs for the active and passive SD operation in Abdelouhab et al. (2010), agreeing with the experimental results found here. Although for the pilot house the difference between the behaviour of active and passive SD is significantly different.

According to the results, the permeability characterisation of the aggregate layer under the slab seems to be different from the results of the active and passive SD operation. But it should be taken into account that the extraction airflow rate during the active SD is measured in the exhaust pipe right under the mechanical fan. As for the passive SD, the wind velocity is recorded at the experimental house rooftop and it can differ from the effective extraction airflow at the exhausting pipe.

Hung et al. (2019) studied the permeability characteristics of Irish aggregates used in construction for the aggregate layer below the slab in relation to its depressurisation ability for radon mitigation. Among other results, the work presents pressure difference generated as function of the extraction airflow for different Irish aggregate materials thickness. Although the experiment setup is different, results can be comparable to the outcomes of the monitoring study at the pilot house (see Figure 10).

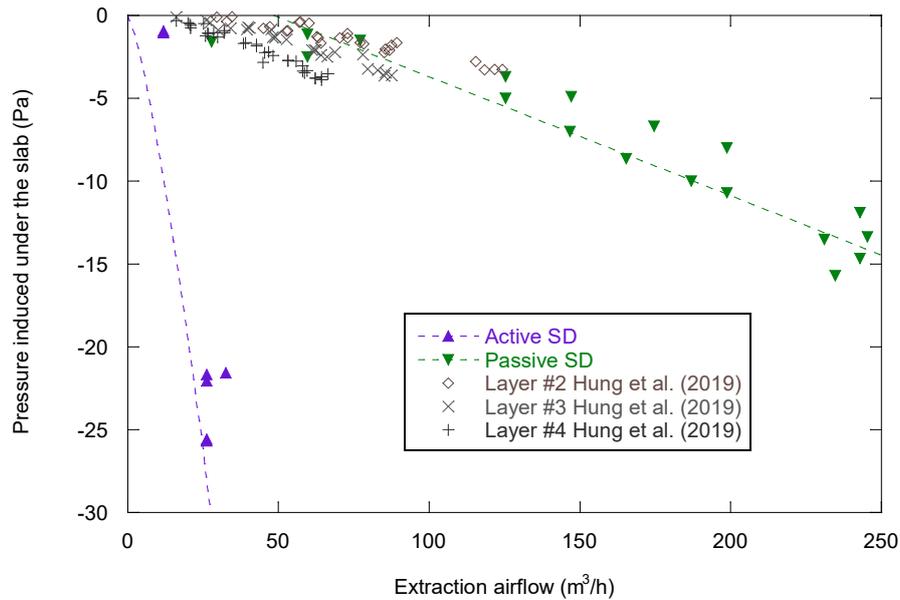

Figure 10. Pressure induced under the slab against extraction airflow through the central exhaust pipe, along with some results from the work of Hung et al. (2019) for the pressure difference obtained through different layers of T2 Perm aggregate material as function of the extraction airflow. Dots represent the experimental data for the active SD in purple and passive SD in green, and the results from Hung et al. (2019) in grey. The dashed lines are the trends obtained for the experimental data.

From Figure 10 it can be seen that at extraction airflow rates below 100 m$^3$/h, the experimental data for the passive SD testing at the pilot house matches the results from Hung et al. (2019) for the permeability study of the Irish T2 Perm aggregate material. This means that the permeability characterisation of the aggregate layer under the slab in the pilot house, based on the passive SD testing results, is similar to a layer of 30 to 60 cm of Irish T2 Perm aggregate material. This result also agrees with the outcomes from previous work published on the benchmarking of granular fill materials in the European context (Fuente et al. 2019).

*4.2 Radon reduction*

The radon reductions obtained for the different testing phases are summarised in Table 2, relative to the average radon concentration in the basement and ground floor calculated from the closed testing phases. It should be emphasised that the outdoor radon concentration in the area surrounding the pilot house is very high, approximately 300 Bq/m$^3$, while the average outdoor radon concentration globally is between 5 – 15 Bq/m$^3$ (WHO, 2016).

In all cases the radon reductions obtained are over 85%, and the highest reduction is found for the testing phase 4 reaching a radon concentration in the ground floor of 328 Bq/m$^3$, which is comparable to the outdoor radon concentration at the site, and a radon concentration of 662 Bq/m$^3$ in the basement. During phase 4 the mechanical fan was tested varying the extraction airflow and up to the highest power permitted (80 W).

Table 2. Radon concentration $C_{Rn}$ found in the basement and ground floor for the SD testing phases indicated and radon reduction, respect to the average radon levels for the closed periods.

|  | $C_{Rn}$ (Bq/m³) | | Radon reduction | |
|---|---|---|---|---|
|  | Basement | Ground Floor | Basement | Ground Floor |
| Average closed | 54625 | 26421 |  |  |
| Phase 2: passive SD | 7417 | - | 86% | - |
| Phase 4: active SD ($v_{ext}$ = 0 - 3.5 m/s) | 662 | 328 | 99% | 99% |
| Phase 6: active SD ($v_{ext}$ = 1.5 m/s) | 3326 | 3689 | 94% | 86% |
| Phase 8: active SD ($v_{ext}$ = 2 m/s) | 3701 | 2279 | 93% | 91% |

From phase 6 experimental results, it can be highlighted that a 34 W mechanical fan, which is the equivalent power for the extraction airflow used during testing phase 6, is sufficient to reach radon reductions up to 94% in the basement and 86% in the ground floor in a house of these permeability characteristics. In terms of pressure induced under the slab, the average value at the sump recorded for testing phase 6 is -55 Pa.

Although it depends on the atmospherics conditions (mainly wind and temperature) and the occupant behaviour, the typical pressure difference found between indoors and the soil in a dwelling oscillates between 0 to 5 Pa. Thus, the depressurisation system should be designed to induce at least -6 Pa in every point of the slab area (Fowler et al., 1991; Broadhead, 2018; Dumais, 2018). Looking at Figure 8a, it can be observed that such pressure is obtained for extraction velocities below 1 m/s, that correspond to 23 W power of the mechanical fan. Therefore, it could be stated that a 23 W mechanical fan is sufficient to achieve an optimum soil depressurisation reaching radon reductions above 85% for dwellings with similar permeability characteristics to the experimental house studied here.

## 5. Conclusions

A monitoring study was conducted in a pilot house with high radon levels to investigate the ability and efficiency of radon mitigation by soil depressurisation (SD) both active and passive. The study was motivated by the need to quantify SD effectiveness in terms of pressure field extension and also quantify significant radon reductions.

The testing plan consisted of different testing phases, alternating closed periods of the house to foster radon accumulation in the experimental house with SD operation phases. The variables were monitored over the different testing periods were radon concentration in the basement and ground floor of the house, pressure field extension under the slab and atmospheric parameters such as wind velocity, outdoor temperature, atmospheric pressure, relative humidity percentage and accumulated rain.

Radon concentration behaviour was analysed for the closed testing periods, an average radon concentration of 55 kBq/m$^3$ was found in the basement, while for the ground floor there is an average radon concentration of 26 kBq/m$^3$. Atmospheric variables influence on the radon behaviour in the house was also studied, finding significant negative correlations between atmospheric pressure and wind velocity with the radon concentration in the house, and a positive correlation with the temperature.

From the analysis of the pressure distribution under the slab it was proven that the pressure drop with distance from the suction point of the SD system is linear with the depressurisation generated under the slab. Still, it was found that the distribution of the pressure under the slab in the pilot house is quite homogeneous, not exceeding 1 Pa/m pressure drop with distance for the highest depressurisation generated under the slab by active SD.

Results from the passive SD operation show a lower and discontinuous depressurisation induced at the sump by the wind velocity compared to depressurisation generated because of the active SD operation using a mechanical fan. However, based on the results analysis for the passive SD operation it was found that the permeability characterisation of the pilot house agrees with previous works published on the characterisation of granular fill materials for radon soil depressurisation systems across Europe.

Finally, radon reductions in excess of 85% were achieved for the different testing phases in all cases. Based on the radon reduction results associated with the depressurisation generated under the slab as function of the extraction airflow for the active SD conditions considered, it is found that 23 W power for a mechanical fan is sufficient to achieve an optimum soil depressurisation reaching radon reductions above 85%.

To summarise, the case study presented contributes to the specification for optimum soil depressurisation systems performance and the findings encountered have applicability within similar building type dwellings with comparable aggregate layer permeability characteristics within the European context.

**Authors' contributions**

Mark Foley, Jamie Goggins, Carlos Sainz and Marta Fuente contributed to the initial research plan and the conception of the monitoring study. Marta Fuente, Carlos Sainz, Ismael Fuente and Daniel Rábago also evaluated and reviewed the testing plan. The experimental development of the study was conducted by Marta Fuente and Daniel Rábago with the help from the radon group technical staff (Enrique Fernández, Jorge Quindós and Luis Quindós) and the support of ENUSA Industrias Avanzadas S.A. Marta Fuente processed the experimental data, analysed the results with the help from Daniel Rábago and wrote the manuscript. Mark Foley, Jamie Goggins, Ismael Fuente and Carlos Sainz reviewed and edited the manuscript for the final version. All authors read and approved the final manuscript.

**Acknowledgments**

This research was supported by the OPTI-SDS project carried out at the National University of Ireland Galway in collaboration with University of Cantabria, Spain. The OPTI-SDS project is funded by the Irish Environmental Protection Agency (EPA) [project code 2015-HW-MS-5].

The work was performed at the ENUSA Industrias Avanzadas S.A. facilities in Spain, thanks to the collaboration with the radon group at University of Cantabria. The authors would like to acknowledge ENUSA Industrias Avanzadas S.A. for the support and help while conducting the experiment and the technical staff of the radon group for their work on the experimental development of the study.